\documentclass[12pt,a4paper]{article}
\usepackage{amsmath}
\usepackage[font={footnotesize,it}]{caption}
\usepackage[top=1.21in, bottom=1.21in, left=1.2in, right=1.2in]{geometry}

\DeclareCaptionStyle{italic}[justification=centering]{labelfont={bf},textfont={it},labelsep=colon}
\usepackage{graphicx,psfrag,epsf}
\usepackage{enumerate}
\usepackage{natbib}
\usepackage{url, setspace}
\usepackage{comment}
\usepackage{multirow}
\usepackage{subcaption}
\usepackage{booktabs, bm, paralist,mathpazo,tikz,todonotes,longtable,microtype,dsfont,rotating} 
\usepackage[pdftex,colorlinks=true]{hyperref}
\definecolor{darkblue}{rgb}{0,0,.6}
\hypersetup{citecolor=darkblue,linkcolor=darkblue,urlcolor=darkblue}

\newcommand{\blind}{0}

\newcommand{\X}{\mathcal{X}}

\newsavebox\CBox
\def\textBF#1{\sbox\CBox{#1}\resizebox{\wd\CBox}{\ht\CBox}{\textbf{#1}}}

\date{\today}

\begin{document}

\def\spacingset#1{\renewcommand{\baselinestretch}
{#1}\small\normalsize} \spacingset{1}

\if0\blind
{
  \title{\bf Functional time series forecasting of extreme values}
  \author{Han Lin Shang\thanks{Postal address: Department of Actuarial Studies and Business Analytics, Level 7, 4 Eastern Road, Macquarie University, Sydney, NSW 2109, Australia; Telephone: +61(2) 9850 4689; Email: hanlin.shang@mq.edu.au; ORCID: \url{https://orcid.org/0000-0003-1769-6430}} 
  \hspace{.2cm}\\
    Department of Actuarial Studies and Business Analytics \\
    Macquarie University \\
    Research School of Finance, Actuarial Studies and Statistics \\
    Australian National University \\
    \\
    Ruofan Xu \\
    Department of Econometrics and Business Statistics \\
    Monash University \\
        Research School of Finance, Actuarial Studies and Statistics \\
    Australian National University
}
  \maketitle
} \fi

\if1\blind
{
   \title{\bf Functional time series forecasting of extreme values}
   \author{}
   \maketitle
} \fi

\bigskip

\begin{abstract}
We consider forecasting functional time series of extreme values within a generalised extreme value distribution (GEV). The GEV distribution can be characterised using the three parameters (location, scale and shape). As a result, the forecasts of the GEV density can be accomplished by forecasting these three latent parameters. Depending on the underlying data structure, some of the three parameters can either be modelled as scalars or functions. We provide two forecasting algorithms to model and forecast these parameters. To assess the forecast uncertainty, we apply a sieve bootstrap method to construct pointwise and simultaneous prediction intervals of the forecasted extreme values. Illustrated by a daily maximum temperature dataset, we demonstrate the advantages of modelling these parameters as functions. Further, the finite-sample performance of our methods is quantified using several Monte-Carlo simulated data under a range of scenarios.
\\

\end{abstract}

\newpage
\spacingset{1.49}

\section{Introduction}\label{S1}

For planning government infrastructure, forecasting extreme values is extremely useful in many scientific disciplines, including but not limited to, Economics \citep[e.g.,][]{calabrese2013modelling}, Hydrology \citep[e.g.,][]{tyralis2019explain} and Meteorology \citep[e.g.,][]{de2011generalized}. Motivated by the need to forecast the extreme temperatures, we propose a functional time series method to forecast the GEV density with an application to Sydney daily maximum temperatures from 1900 to 2019.

Monitoring the temperature extremes is of critical importance in the study of climate impact, as the extreme weather events generally have large negative impacts on both social and environmental systems \citep{field2012managing}. Temperature extremes are changing over time. Over the last two decades, numerous studies have identified changes in temperature extremes both in observations \citep[e.g.,][]{naveau2014non} and in general circulation model simulations of future climate \citep[e.g.,][]{kharin2013changes}. Therefore, it is crucial not only to identify but also forecast the temperature extremes.
 
The definition of ``extreme" may vary in the literature across multiple disciplines. However, from a statistical point of view, one common definition is based on occurrences in the far tail of the distribution of the quantity of interest. In statistical literature, extreme value theory provides a mathematical framework for studying these far tails \citep{fisher1928limiting}. One common approach making use of the extreme value theory is based on the “block extremes", the maxima (minima) of some climate variable over given blocks of time. Under certain regularity conditions, the magnitudes of extremes over sufficiently long blocks approximately follow a GEV distribution.

The GEV distribution is a well-developed distribution in statistical literature \citep[e.g., see][Chapter 1 for details]{de2007extreme}. In general, the GEV distribution can be fully characterised using only three parameters: the location, scale and shape parameters. As a result, forecasting GEV can be reduced to the forecasting of these three latent parameters. However, the classical forecasting methods for multivariate time series are not directly applicable, since the scale parameter is constrained to be non-negative. In contrast, the location and shape parameters can take arbitrary real values. A natural way to deal with such a constraint is to perform an invertible transformation, for example, the Box-Cox transformation \citep{box1964analysis} and the Yeo-Johnson transformation \citep{YJ00}, that maps the shape parameter onto the real line first and then make the prediction before inverting back.

In the field of meteorology, much attention has been focused on the study of annual extremes under a GEV distribution with three scalar parameters, since blocks of one year are common and sensible \citep[see, e.g.,][]{huang2015estimating, cooley2009extreme}, and annual extremes are generally weakly dependent and stationary. In contrast, little attention has been paid to the study of daily extremes, which is of great value as it gives a detailed guideline of how extremes change among four seasons in a year and across years. Because of the inherent seasonality, it is not reasonable to model the daily maximum temperature within a year using a GEV distribution with \textit{constant} scalar parameters. To this end, a generalised additive extreme value (GAEV) model, which is a GEV model with parameters modelled with generalised additive models, is often considered \citep[see, e.g.,][]{yee2007vector,chavez2005generalized}. To capture the daily variation in the annual extreme temperature, time can be treated as a covariate. In the statistical literature, GEV with covariate effects has been studied using different approaches. For example, \cite{eastoe2009modelling} transformed the sample to remove the effects of covariates prior to extreme value analysis; \cite{jonathan2013joint} considered a conditional extremes model to incorporate covariate effects in all of the threshold selection, marginal and dependence modelling. In this paper, we follow \cite{davison1990models} and parameterise extreme value model parameters in terms of the covariate time, so that the temporal dependency within the year can be preserved. It is worth noting that more sophisticated techniques have been developed for extreme value analysis of time series. \cite{chavez2012modelling} provided a detailed review of the recent development in the time series's extremes. One of the main steam is to ``decluster'' the time series into approximately independent blocks given covariates. A basis reference for this approach is \cite{ferro2003inference}, which proposed an automatic declustering scheme. Another approach is to use the physics of the problem to create a hierarchical model involving extreme value analysis of the ``peaks'' of whole heatwaves only, and then different
models to characterise the evolution of a heatwave relative to its peak. This Bayesian modelling approach is often used in an oceanographic setting to model the evolution of severe ocean storms \citep[see, e.g.,][]{randell2016bayesian,tendijck2019model}.

We focus on the one-year-ahead forecast of daily maximum temperatures. We aim to model and forecast the cumulative distribution function (CDF) of extreme values and produce point and interval forecasts of extreme values associated with certain quantiles of the CDF. One natural way is to construct a sliced functional time series by breaking the long daily temperature data set into annual temperature curves with daily temperatures being the finite realisations on each curve. That is, the observed data are of the form $\{\X_t(\tau_j),t = 1,\dots,T, j = 1,\dots,366\}$, where $\X_t(\tau_j)$ is the maximum temperature on the $j^{\text{th}}$ day in the $t^{\text{th}}$ year. Since the annual data contains 365 observations in common years and 366 observations in leap years, we treat the observation on the missing day, February 29, in common years as a missing value and use the average of the adjacent observations as its estimated value. Then, the yearly objects are observed on a common grid of 366 days. By treating the univariate time series as a functional time series, the one-year-ahead daily maximum temperatures can be predicted as a whole. For each year, a GAEV model can be fitted to the daily maximum temperature curve. With the estimated parameters in the GAEV model, a time series forecasting algorithm is applied to forecast the future parameters in the GAEV model.

The rest of the paper is organised as follows. In Section~\ref{S2}, We first introduce the background of the GEV distribution and then provide a forecasting algorithm for the GEV distribution with scalar parameters. This scalar GEV method will be considered as one of the benchmarks and later compared with the GAEV model using a simulation study and an empirical data analysis. In Section~\ref{S3}, we introduce our forecasting method for the GAEV model and compare the forecasting method with a number of existing methods using a daily maximum temperature data set in Section~\ref{S4}. To measure forecast uncertainty, we apply a sieve bootstrap method of \cite{Efstathios2020pi} to construct pointwise and simultaneous prediction intervals of the forecasted extreme values. Further, we assess forecast accuracy using the simulated data sets in Section~\ref{S.simulation}. Finally, we summarise the research findings, along with some ideas on how the methodology can be further extended in Section~\ref{S5}.

\section{Forecasting extremes using GEV with scalar parameters}\label{S2}

We consider a functional time series $\{\X_{t}(\tau),t = 1, 2, \dots,T, \tau \in \mathcal{I}\}$, where $T$ denotes the sample size, $\mathcal{I} \subset \mathbb{R}$ is a compact interval on the real line, and the observed data is $\X_{t}(\tau_j)$ for $j=1,2,...,J$, where $J$ denotes the number of discrete data points in a curve. In the maximum temperature data, $\mathcal{I} = [1,366]$ and $J=366$.  

In this section, we assume that $\X_t(\cdot)$ follows a GEV distribution with time-specific scalar parameters, i.e. $\X_t(\cdot)\sim \text{GEV}(\mu_t,\sigma_t,\xi_t)$. We consider the simplest case that $(\mu_t,\sigma_t,\xi_t)$ are scalar parameters with respect to $\tau$. In Section~\ref{S3}, these latent parameters are considered to be generalised additive models of a functional variable $\tau$. When the GEV parameters are function-valued, each GEV parameter function can be well approximated by cubic regression splines. These spline basis functions are able to model temporal dependence exhibited in each GEV parameter.

As the GEV densities are not directly observable, we first estimate the time-varying parameters in the GEV densities and then model these estimated GEV parameters $\{(\mu_t,\sigma_t,\xi_t),t = 1,\dots,T\}$ to forecast the future GEV density. In Section~\ref{S2.1}, we present a brief overview of the GEV distribution and the corresponding parameter estimation method. Then, we introduce our time series forecasting method for the time-varying GEV scalar parameters in Section~\ref{S2.2}. In Section~\ref{S2.3}, a complete forecasting algorithm is summarised.

\subsection{The GEV distribution and its parameter estimation}\label{S2.1}

The GEV distribution arises in many cases of natural data. When considering the daily maximum temperature extremes within a year, let $(Y_{j,1},\dots,Y_{j,K})$ be the temperatures recorded densely within day $j$, so that $M_j = \text{max}(Y_{j,1},\dots,Y_{j,K})$ is the daily maximum temperature. The extreme value theorem states that if random variables $(Y_{j,1},\dots,Y_{j,K})$ are independent and identically distributed (i.i.d.), with “block length” $K$ sufficiently large, the maxima $M_j$ converge to a GEV distribution as $j \rightarrow \infty$ \citep{fisher1928limiting, coles2001introduction}. Moreover, there are theoretical justification that the i.i.d condition on $(Y_{j,k})_{k=1}^{K}$ can be relaxed to weakly dependent stationary time series \citep[see, e.g.,][]{hsing1991tail, einmahl2016statistics}, which is also the case we consider here. Note that $M_j$ corresponds to one daily observation $\X_t(\tau_j)$ on the sliced functional time series $\X_t(\cdot)$ in day $j$ at year $t$.

According to \citet{coles2001introduction}, the probability density function of the GEV distribution is defined as
\begin{equation}
\label{GEV.pdf}
    f(x|\mu,\sigma,\xi)
   = \begin{cases}
   \frac{1}{\sigma}\big(1+\xi \frac{x-\mu}{\sigma}\big)^{-1 / \xi-1} \exp \left[-(1+\xi \frac{x-\mu}{\sigma})^{-1 / \xi}\right], & \xi \neq 0, \\
   \frac{1}{\sigma}\exp (-\frac{x-\mu}{\sigma})\exp \big[-\exp (-\frac{x-\mu}{\sigma})\big], & \xi=0,
   \end{cases}
\end{equation}
where $u_{+} = \text{max}(0,u)$, $\mu \in \mathbb{R}$ is the location parameter, $\sigma > 0$ is the scale parameter, and $\xi \in \mathbb{R}$ is the shape parameter. When $\xi >0$, the GEV distribution has a finite upper tail for the shape parameter. In contrast, there is no upper bound when $\xi \leq 0$. 

Because of its relation with quantile function, we choose to work with the CDF. The CDF for a random variable $X \sim \text{GEV}(\mu,\sigma,\xi)$ is given by
\begin{equation*}
\label{GEV.cdf}
    F(x|\mu,\sigma,\xi)
   = \begin{cases}
   \exp \left[-(1+\xi \frac{x-\mu}{\sigma})_{+}^{-1 / \xi}\right], & \xi \neq 0, \\
   \exp \Big[-\exp \big(-\frac{x-\mu}{\sigma} \big)\Big], & \xi=0.
   \end{cases}
\end{equation*}

Since the CDF is invertible, the quantile function for the GEV distribution has an explicit expression, namely, for any probability $p \in [0,1]$, the quantile is given by
\begin{equation*}
\label{GEV.quantile}
    Q(p|\mu,\sigma,\xi)
   = \begin{cases}
   \mu + \frac{\sigma \big[(-\ln(p))^{-\xi}-1\big]}{\xi}, & \xi > 0, p\in [0,1);\ \xi < 0, p\in (0,1], \\
   \mu - \sigma \ln[-\ln\big(p)], & \xi=0, p \in (0,1).
   \end{cases}
\end{equation*}

In the statistical literature, several methods have been used to estimate the parameters of the GEV distribution, for example, the method of moments \citep[e.g.,][]{christopeit1994estimating}, the Bayesian method \citep[e.g.,][]{coles2005bayesian} and the maximum likelihood method \citep[e.g.,][]{smith1987comparison}. Among them, the maximum likelihood method is the most popular one, because it allows additional components to the fitting of covariates, such as trends and cycles \citep{katz2002statistics}. Therefore, we use the maximum likelihood method to estimate the parameters of the GEV distribution as follows. 

If $M_1,\dots,M_J \overset{\text{i.i.d}}{\sim} \text{GEV}(\mu,\sigma,\xi)$ with $\xi \neq 0$, the likelihood function $\mathcal{L}$ is given by
\begin{align*}
 & \mathcal{L}( \mu, \sigma,\xi ; M) \\
 &= \prod_{i=1}^{J} \frac{1}{\sigma}\left[1+\xi\frac{m_{i}-\mu}{\sigma}\right]^{-\frac{1}{\xi}-1}
\exp \left\{-\left[1+\xi\frac{m_{i}-\mu}{\sigma}\right]^{-\frac{1}{\xi}}\right\}\\
& = \sigma^{-j} \exp \left\{\sum_{i=1}^{J} -\left[1+\xi\frac{m_{i}-\mu}{\sigma}\right]^{-\frac{1}{\xi}}\right\}
\prod_{i=1}^{J} \left[1+\xi\frac{m_{i}-\mu}{\sigma}\right]^{-\frac{1 }{\xi}-1},
\end{align*}

By taking the natural-log transformation, the log-likelihood is 
\begin{equation}
l(\mu, \sigma,\xi ; M)=-J \ln \sigma-\left(1+\frac{1}{\xi}\right) \sum_{i=1}^{J} \ln \left(1+\xi\frac{m_{i}-\mu}{\sigma}\right) -\sum_{i=1}^{J}\left[1+\xi\frac{m_{i}-\mu}{\sigma}\right]^{-\frac{1}{\xi}}. \label{GEV.MLE}
\end{equation}

The maximum likelihood estimation (MLE) is then obtained by maximising~\eqref{GEV.MLE} under the constraints that
$1+\xi \frac{x-\mu}{\sigma} >0$ and $\sigma>0$. There is no analytic solution to this optimisation problem no matter whether the constraints are met. However, the numerical solution $(\widehat{\mu}$, $\widehat{\sigma}, \widehat{\xi})$ can be obtained using the \texttt{fgev} function of the \texttt{evd} package \citep{Stephenson02}.

\subsection{Forecasting approach for the time-varying GEV parameters}\label{S2.2}

Since the daily maximum follows a GEV distribution within each year, the GEV parameters (mean: $\mu_t$, scale: $\sigma_t$, shape: $\xi_t$) are estimated for each year $t$ independently, and subscript $t$ aims to distinguish the year. Furthermore, $\{(\mu_t,\sigma_t,\xi_t), t=1, \dots,T\}$ forms a vector-valued time series with the constraint that $\sigma_t$ is positive for all $t=1, \dots,T$. Classic time series models, which have no constraint on the domain, can not be directly applied to this vector-valued time series. To this end, we first map $\sigma_t$ onto the real line with an invertible transformation, such as the Box-Cox transformation. As the transformed time series may not be stationary, additional transformation, such as log-transformation, de-trending or differencing, may also be necessary to obtain an approximately stationary vector-valued time series $\boldsymbol{\theta}_t$. A stationary time series model can be fitted and then the $h$-step-ahead forecasts, $\widehat{\boldsymbol{\theta}}_{T+h}$, can be obtained accordingly. One simplest model in this case would be the vector autoregressive model, whose order can be selected via the corrected Akaike information criterion (AICC) \citep[see, e.g.,][]{hurvich1993corrected}. Finally, using the components of $\widehat{\boldsymbol{\theta}}_{T+h}$, we can compute the predicted parameters $(\widehat{\mu}_{T+h},\widehat{\sigma}_{T+h},\widehat{\xi}_{T+h})$ through an inverse transformation of $\widehat{\boldsymbol{\theta}}_{T+h}$. 

\subsection{An algorithm for the GEV density prediction}\label{S2.3}

The time series forecasting method for the GEV density can be summarised as follows.
\begin{enumerate}
    \item[1)] For each $t=1,2,\dots,T$, maximise~\eqref{GEV.MLE} under the constraints that
$1+\xi \frac{x-\mu}{\sigma} >0$ and $\sigma>0$ to obtain the MLEs $\widehat{\mu}_t$, $\widehat{\sigma}_t$ and $\widehat{\xi}_t$.
    \item[2)] Transform the estimated parameters $\{(\mu_t,\sigma_t,\xi_t)\}$ to obtain a stationary unconstrained vector time series $\{\boldsymbol{\theta}_t\}$.
    \item[3)] Fit an appropriate time series model to $\{\boldsymbol{\theta}_t\}$.
    \item[4)] Compute the $h$-step-ahead forecast $\widehat{\boldsymbol{\theta}}_{T+h}$ for $h\geq 1$.
    \item[5)] Using $\widehat{\boldsymbol{\theta}}_{T+h}$ to compute the predicted parameters $(\widehat{\mu}_{T+h},\widehat{\sigma}_{T+h},\widehat{\xi}_{T+h})$ via back-transformation.
    \item[6)] Obtain the predicted GEV densities $\widehat{f}_{T+h}(\tau) = f(u|\widehat{\mu}_{T+h},\widehat{\sigma}_{T+h},\widehat{\xi}_{T+h})$ according to~\eqref{GEV.pdf}.
\end{enumerate}

\section{Forecasting extremes with a GAEV model}\label{S3}

When the GEV parameters are modelled as scalars, it has great limitation and is often unrealistic, as observations on a curve are rarely i.i.d. Also, it is not reasonable to assume that all realisations on each curve follow the same GEV distribution. To capture the temporal dynamics in the GEV distribution along a curve, a GAEV model is considered. The GAEV is a GEV model with parameters characterised by generalised additive models \citep[see, e.g.,][]{gilli2006application,chavez2005generalized}. To be more specific, 
\begin{equation}
\label{GAEVM}
\X_{t}(\tau) \sim \text{GEV}[\mu_{t}(\tau),\sigma_t(\tau),\xi_t(\tau)], \quad \forall t \in {1,\dots,T}, \quad \tau \in \mathcal{I},
\end{equation}
where $\sigma_t(\tau)$ is a non-negative function since the scale parameter is required to be non-negative.

Let $\eta^{*}_t$ denote any of the three functional parameters, namely $\eta^{\mu}_t(\tau) := \mu_t(\tau)$, $\eta^{\sigma}_t(\tau) := \ln[\sigma_t(\tau)]$ and $\eta^{\xi}_t(\tau) := \xi_t(\tau)$, the generalised additive model for $\eta^{*}_t$ can be represented using a basis expansion
\begin{equation}
\label{etaGAM}
\eta^{*}_t(\tau) = \beta^{*}_{t,0} + \sum_{i=1}^{d^{*}}\beta^{*}_{t,i}b^{*}_{i}(\tau),
\end{equation}
where $d^{*}$ is a pre-determined positive integer, $\{b^{*}_{i}(\tau),i=1,\dots,d^{*}\}$ are the pre-determined basis functions and $\{\beta^{*}_{t,i},i=0,\dots,d^{*}\}$ are the time-varying coefficients. In this paper, we use the cubic regression spline basis as the basis functions.

For the GAEV model in~\eqref{GAEVM}, the GEV parameters of the generalised additive model (GAM) form are estimated by maximising a penalised likelihood. Such estimation has been implemented in the R package \texttt{evgam} \citep{youngman2020evgam}. 

As the GEV functional parameters can be fully characterised by the basis coefficients $\{\beta^{*}_{t,i},i=0,\dots,d^{*}\}$, the forecast for the GAEV model can be performed through the prediction of those basis coefficients, which forms a vector time series $\boldsymbol{\beta}_t=[\beta_{t,i}^{\mu},\beta_{t,j}^{\sigma},\beta_{t,l}^{\xi},i=0,...,d^{\mu},j=1,...,d^{\sigma},k=1,...,d^{\xi}]$. We fit a vector autoregressive (VAR) model to $\{\boldsymbol{\beta}_t,t=1,...,T\}$ and obtain the $h$-step-ahead time series prediction $\{\widehat{\beta}^{*}_{T+h,i},i=0,\dots,d^{*}\}$, then the forecast $\widehat{\eta}^{*}_{T+h}(\tau)$ can be computed via~\eqref{etaGAM} with $\beta^{*}_{t,i}$ replaced by $\widehat{\beta}^{*}_{T+h,i}$ and estimated basis functions. Through an inverse log-transformation, we obtain $\widehat{\eta}^{\sigma}_{T+h}(\cdot)$. The $h$-step-ahead GAEV model prediction is obtained as $\widehat{\X}_{T+h}(\tau) \sim \text{GEV}\{\widehat{\eta}^{\mu}_{T+h}(\tau),\text{exp}[\widehat{\eta}^{\sigma}_{T+h}(\tau)],\widehat{\eta}^{\xi}_{T+h}(\tau)\big\}$. 

As the dimension $d^*$ are unknown, in this paper, we apply the leave-one-out cross-validation technique to determine the dimension $d^{\mu}$, $d^{\sigma}$ and $d^{\xi}$ jointly. We set the upper bound of $d^*$ to be 10 (and the lower bound is $3$ for cubic regression spline), then for all possible combinations of $(d^{\mu}, d^{\sigma}, d^{\xi})$, we fit the GAEV model using the first $T-1$ data points, make one-step-ahead forecast $\widehat{\X}_{T}(\tau)$. Since we only observe one single sample for each $\tau$ instead of a complete distribution, we evaluate $\widehat{\X}_{T}(\tau)$ at $50\%$ quantile for each $\tau$, $\widehat{\X}_{T}^{0.5}(\tau)$, and compute the Jensen-Shannon divergence (JSD) between the observed $\X_{T}(\tau)$ and $\widehat{\X}_{T}^{0.5}(\tau)$ (see Eq.~(\ref{JSD})). We choose the set of $(d^{\mu}, d^{\sigma}, d^{\xi})$ resulting in the minimum JSD to be the appropriate dimension parameters to fit the complete data.

\section{Application to daily maximum temperatures in Sydney} \label{S4}

We consider the daily maximum temperature data recorded from a weather station in Sydney from 1900 to 2019. We assume that the daily maximum temperatures in a given year follow a GEV distribution, which is common when studying the temperature data \citep[see also][]{stein2017should, huang2015estimating}. We evaluate the one-year-ahead forecast of the daily maximum temperatures under the GEV model with scalar parameters in Section~\ref{S2} and compare the forecast with the one obtained from the GAEV model in Section~\ref{S3}. As a benchmark, we also consider a GAEV model applied to the univariate time series record. This is a traditional method for forecasting extreme values in the univariate time series literature \citep[see, e.g.,][]{coles2001introduction, mcneil2000estimation}. We first introduce the motivating data set in Section~\ref{S4.1}, then the forecasting approaches in Section~\ref{S4.2}, followed by discussions in Section~\ref{S4.4}. In Section~\ref{S4.5}, we provide a sieve bootstrap 95\% pointwise and simultaneous prediction intervals for one-year-ahead daily maximum temperature extremes at the 99.9\% quantile.

\subsection{Data set}\label{S4.1}

The data set was obtained from the \cite{WB:2019}. The raw data contains the daily maximum temperature from January 1, 1900, to December 31, 2019, collected from the weather station (station number 66062) in the south of Sydney. This data set can be obtained upon request from the corresponding author. Since the temperatures show apparent annual cycle at the location, the univariate daily temperature records are split into 120 yearly records. That is, when considering the maximum temperature extremes, the random variables $\{\X_{t}(\tau_j),t = 1,\dots,T, j =1,\dots,J\}$ are the $j$\textsuperscript{th} daily maximum temperature in year $t$ with $T = 120$ and $J = 366$.

\begin{figure}[!htbp]
   \centering 
    \begin{subfigure}[b]{0.5\textwidth}
                \includegraphics[width=\linewidth]{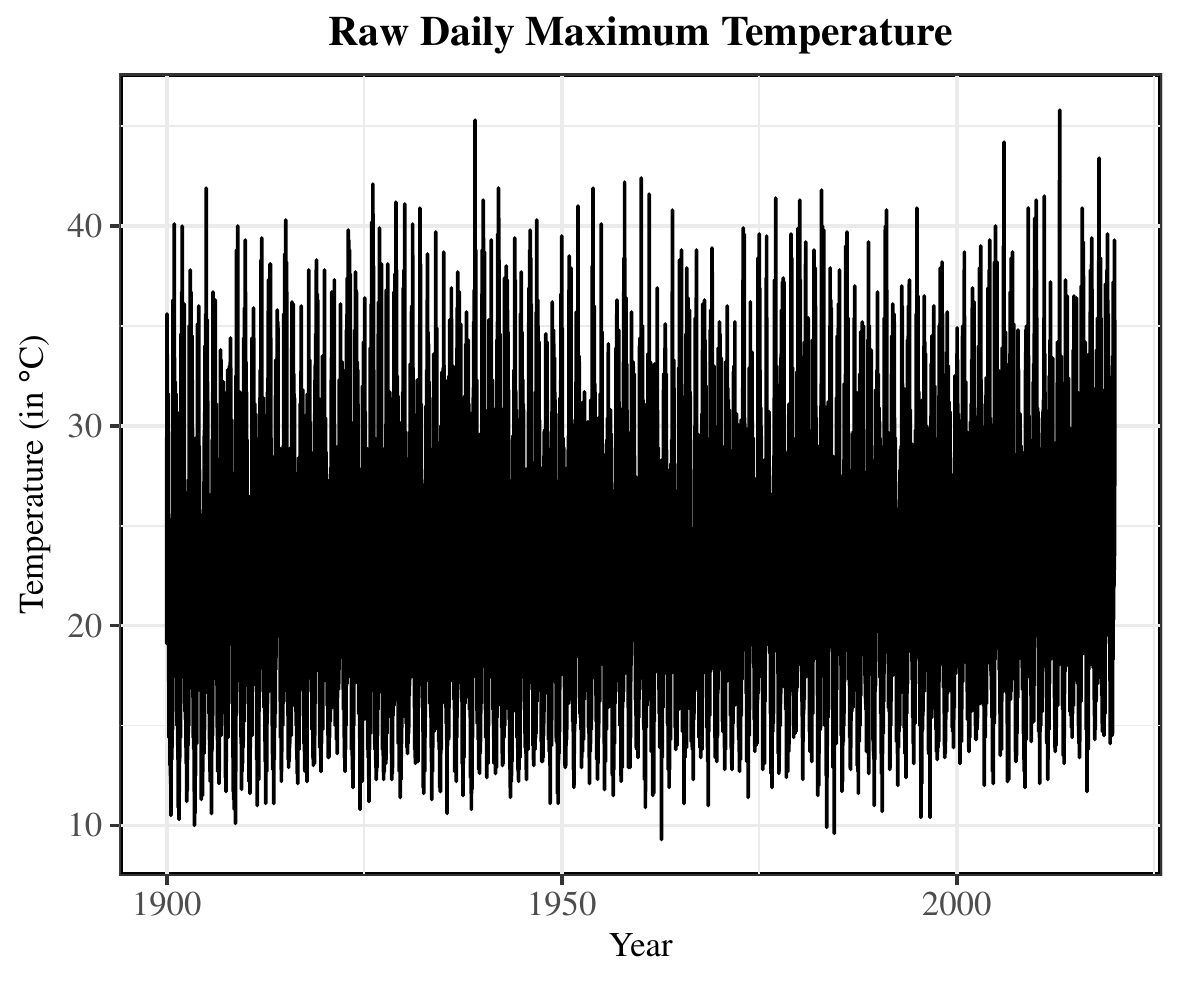}
        \end{subfigure}%
        \begin{subfigure}[b]{0.5\textwidth}
                \includegraphics[width=\linewidth]{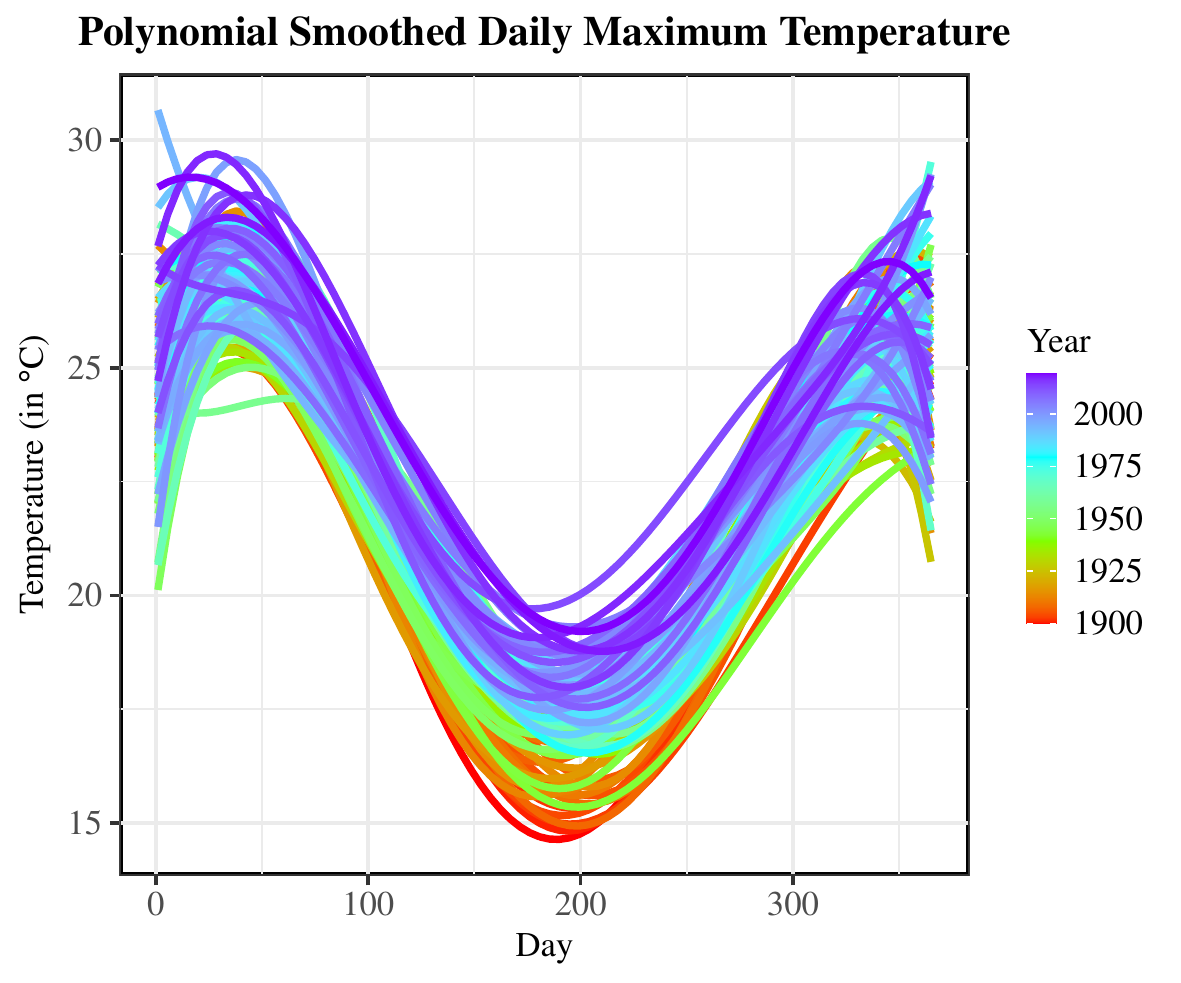}
        \end{subfigure}%
    \caption{The raw and smoothed daily maximum temperatures from 1900 to 2019}
    \label{data}
\end{figure}

Figure~\ref{data} presents a univariate time-series plot for the raw daily temperature from 1900 to 2019 and the rainbow plot of the sliced daily maximum temperature curves. The long time series record on the left panel suggests that the time series is roughly stationary. On the right panel, the seasonal difference in the maximum temperature can be easily spotted along each curve, which implies that a GAVEM may be more appropriate compared to the GEV with scalar parameters.

\subsection{Forecasting methods}\label{S4.2}

In the empirical data analysis, comparisons are made based on the following three GEV models. 

\subsubsection{A GEV model with scalar parameters on functional time series (fGEV)}

We first consider a GEV model with scalar parameters, which ignores the seasonal effect in the daily temperature. The forecasting algorithm is presented in Section~\ref{S2.3}. We make a few specifications regarding steps 2 to 4 in the algorithm: log-transformation will be performed on $\{\widehat{\sigma}_t, t=1,\dots,T\}$ to remove the positive constraint on the scalar parameter. Then one-step-ahead forecasts for the transformed parameters are performed using the VAR model.

\subsubsection{A GAEV model on univariate scalar time series (tsGAEVM)}

To capture the mean shift and seasonality in the GEV distribution, we also consider a na\"{i}ve GAEV model performed on the original daily maximum temperature from January 1, 1990, to December 31, 2019. With such a long univariate time series, the upward trend may dominant the model, consequently resulting in a poor forecast result. Hence, in Section~\ref{S4.4}, we only use the latest one-year data, that is the daily temperature data in the year 2019, to fit the GAEV model. That is $X_t \sim \text{GEV}\big(\mu(t),\sigma(t),\xi\big)$ for $t = 1,\dots, 366$. One-to 366-step-ahead forecasts are performed to obtain the daily maximum temperature forecast in the year 2020. The model fitting and prediction are implemented using the \texttt{evgam} package \citep{youngman2020evgam} in R \citep{Team20}.

\subsubsection{A GAEV model on functional time series (fGAEVM)}

We consider our recommended methodology, GAEV model on functional time series. We assume that $\X_t(\tau) \sim \text{GEV}\big[\mu_t(\tau),\sigma_t(\tau),\xi_t\big]$. Similar to the tsGAEVM mentioned above, we assume that the location and scale parameter follows a GAM, while the shape parameter is a scalar. The GAM for the location and scale parameter aim to capture the mean shift and scale change of the GEV distribution resulting from the seasonal effect. In contrast, the shape parameter is only fitted with intercept for the following two reasons: Firstly, data generally provide little information on the shape parameter. Secondly, estimation is numerically fraught when parameters are allowed to be too flexible \citep{yee2007vector}. 

Having estimated $[\widehat{\mu}_t(\tau),\widehat{\sigma}_t(\tau),\widehat{\xi}_t]$, the one-step-ahead forecasts for the three parameters are obtained jointly by fitting a VAR model to $[\widehat{\mu}_{t}(\tau),\ln(\widehat{\sigma}_{t}(\tau)),\widehat{\xi}_{t}]$ for $t=1,\dots,T$; the order of the VAR model is determined by AICC.

\subsection{Results and Discussion}\label{S4.4}

\begin{figure}[!htbp]
    \centering 
    \begin{subfigure}[b]{0.32\textwidth}
                \includegraphics[width=\linewidth]{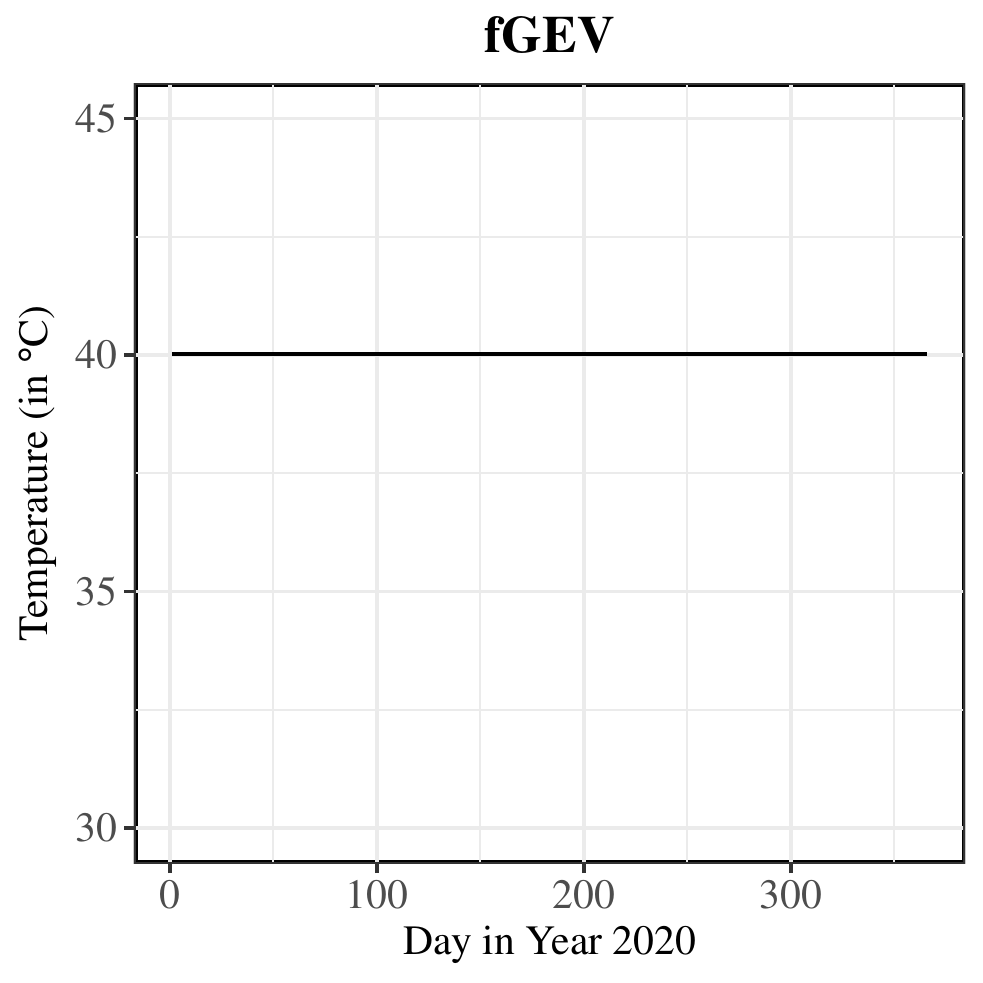}
        \end{subfigure}%
        \begin{subfigure}[b]{0.32\textwidth}
                \includegraphics[width=\linewidth]{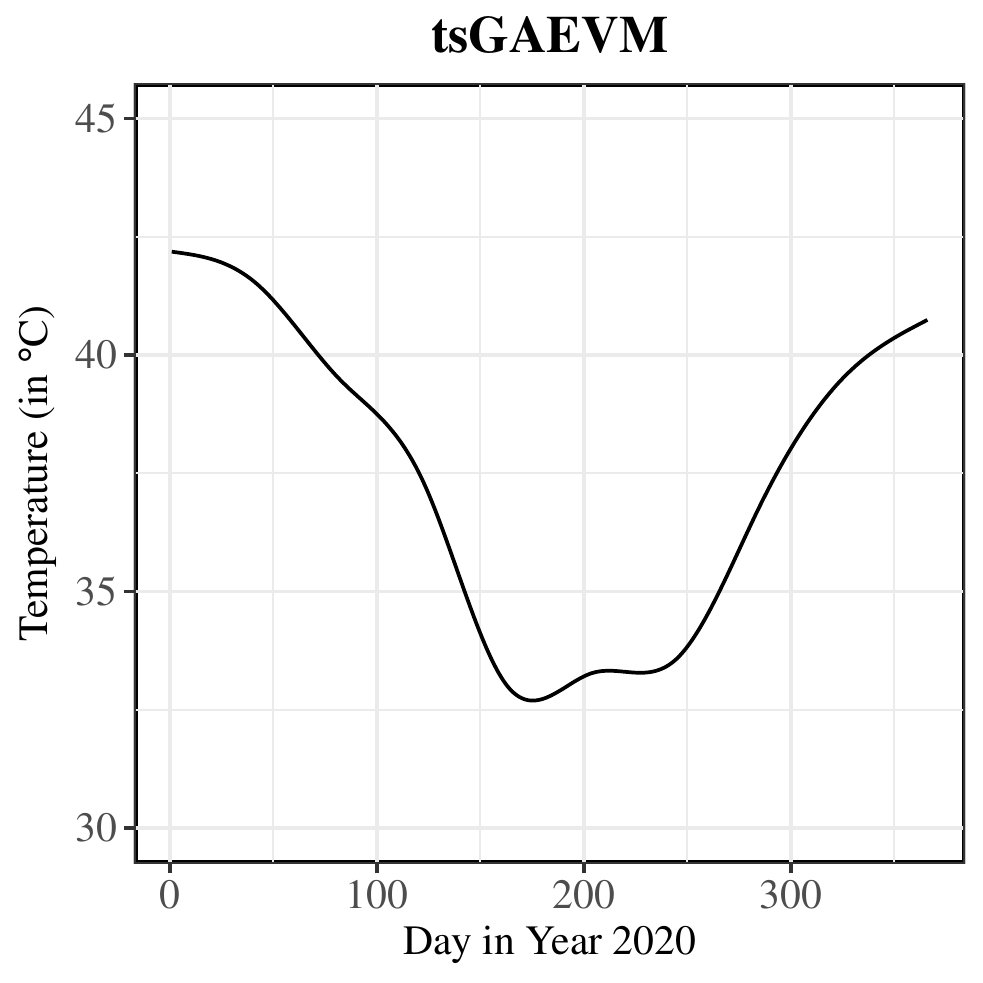}
        \end{subfigure}%
        \begin{subfigure}[b]{0.32\textwidth}
                \includegraphics[width=\linewidth]{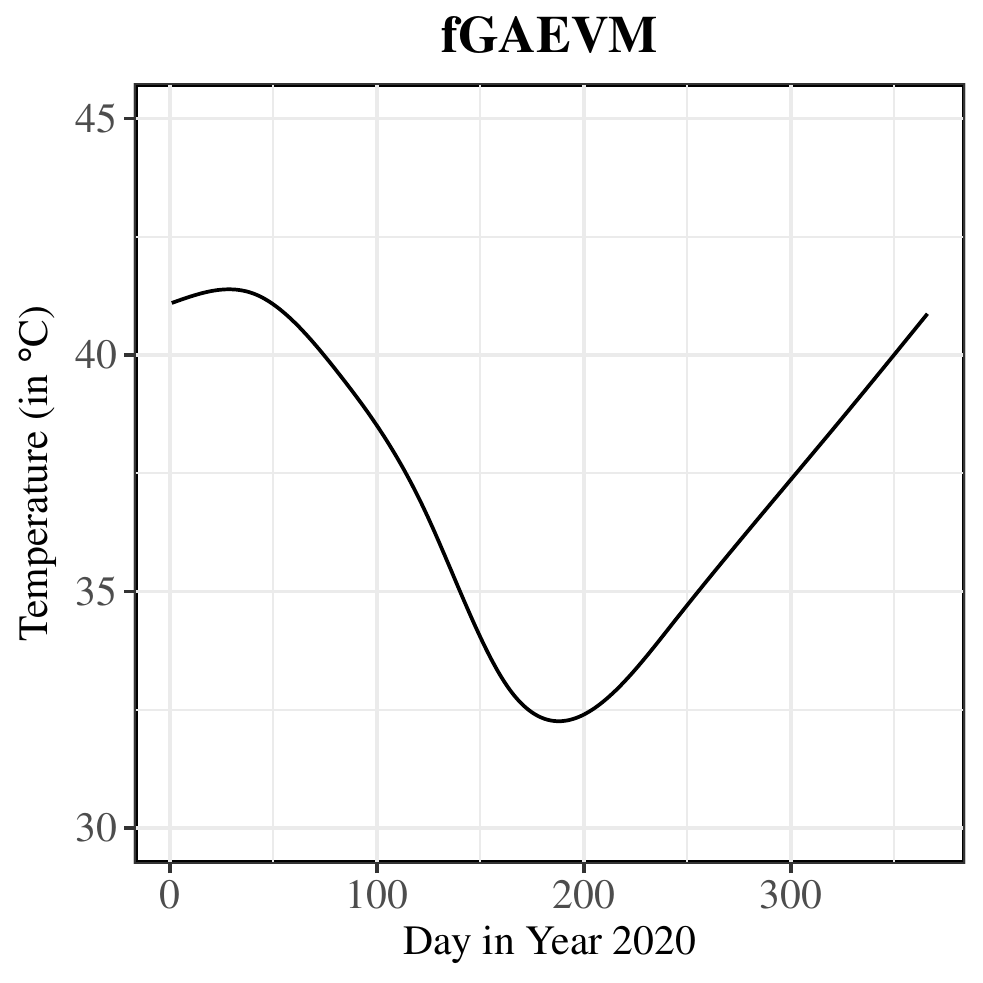}
        \end{subfigure}%
    \caption{The forecast result for the extreme daily maximum temperatures in the Year 2020 at a 99.9\% quantile}
    \label{forecasts}
\end{figure}

As the extreme temperatures are of particular interest in meteorology, Figure~\ref{forecasts} presents the forecast maximum daily temperature curve at a 99.9\% quantile of the GEV distribution. Plots from left to right correspond to the three forecasting models mentioned in section~\ref{S4.2}. The forecasts using the GEV with scalar parameters is a horizontal line, as all temperatures follow the same GEV, which contradicts to the real case. On the other hand, the forecasts using the tsGAEVM and fGAEVM exhibit `U' shape, which reflects the seasonal effect, and hence is more realistic compared to the GEV model. Compared with the tsGAEVM, the smooth forecast using our proposed fGAEVM is more convincing. Although we can not measure the goodness of fit, that wired rough shape in the tsGAEVM is hard to interpret. 

\subsection{Bootstrap prediction interval}\label{S4.5}

\begin{figure}[!htbp]
    \centering 
    \includegraphics[width=0.65\linewidth]{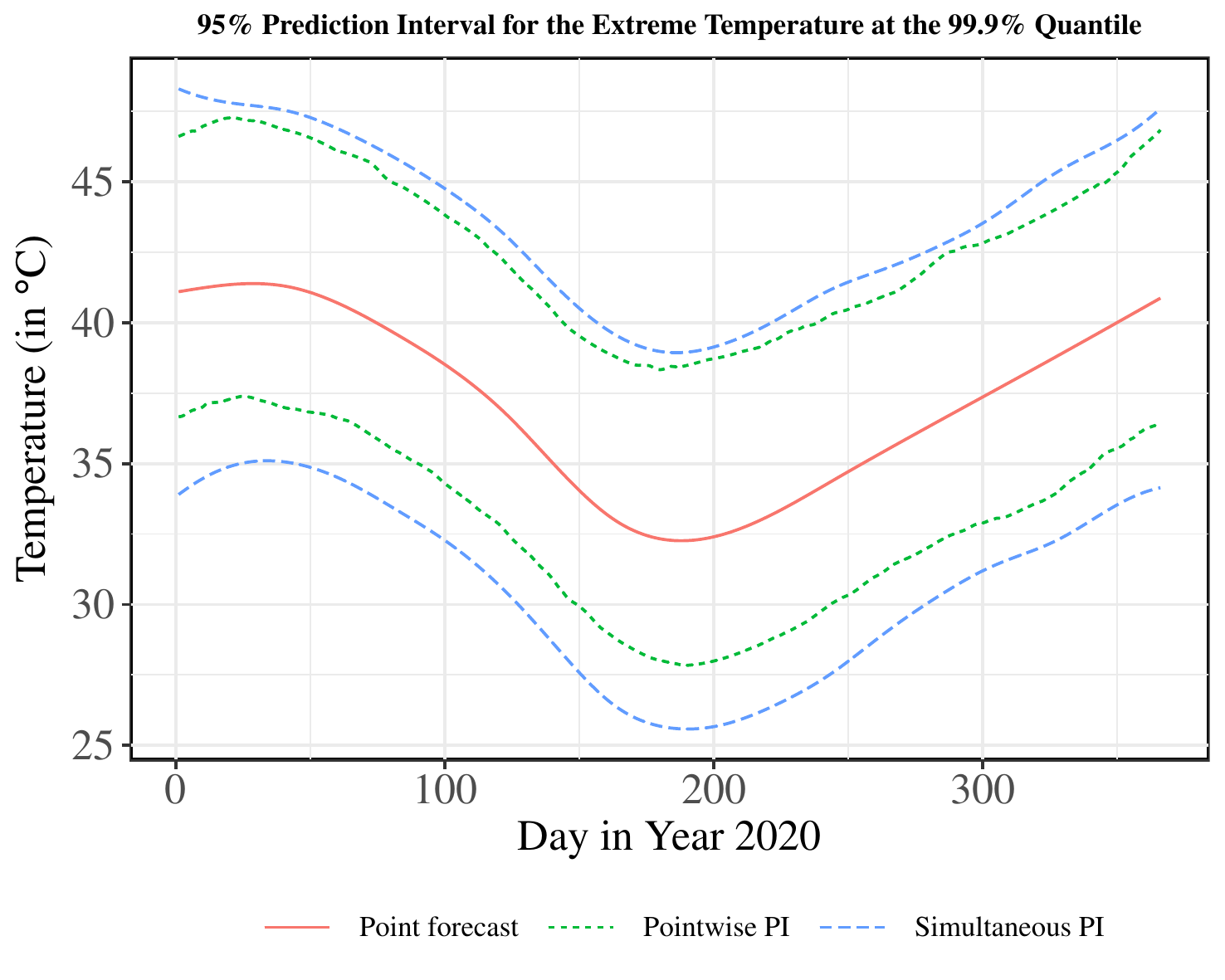}
    \caption{The 95\% pointwise prediction interval and 95\% simultaneous prediction band for the extreme daily maximum temperatures in 2020 at the 99.9\% quantile}
    \label{forecastsPI}
\end{figure}

To quantify the forecast uncertainty associated with the point forecast, we construct $95\%$ pointwise and simultaneous prediction intervals using a sieve bootstrap procedure of \cite{Efstathios2020pi}. The advantage of the sieve bootstrap is its ability to take into account the misspecification error of a forecasting model. Through the sieve bootstrap, we construct $B=1,000$ one-step-ahead bootstrap forecasts, from which we construct the pointwise and simultaneous prediction intervals for the 99.9\% quantile. For the 99.9\% quantile, the 95\% pointwise prediction interval and 95\% simultaneous prediction band are displayed in Figure~\ref{forecastsPI}. 

\section{Simulation study} \label{S.simulation}

The forecast accuracy can not be measured in the real data analysis, since the actual GEV distribution is unknown. To investigate the finite-sample performance, in this section, we utilise a Monte-Carlo simulation study to compare the forecast accuracy for the three GEV models used in the empirical data analysis (see Section~\ref{S4.2}).

\subsection{Data generating processes}\label{S.dgp}

We generate the functional time series $\{\X_t(\tau) \sim \text{GEV}(\mu_t(\tau),\sigma_t(\tau),\xi_t(\tau))$,$t = 1,...,T$, $\tau \in [0,1] \}$ that follow a GAEV model in~\eqref{GAEVM} and~\eqref{etaGAM} and consider three settings regarding the GEV parameters as follows. 

\subsubsection*{Setting 1: All parameters are scalars}

We consider the first setting that all parameters are scalars, that is, 
\begin{align*}
\mu_t(\tau) = \mu_t,\ 
\sigma_t(\tau) = \sigma_t,\ \text{and }
\xi_t(\tau) = \xi_t,
\end{align*}
where all parameters $\mu_t$, $\ln(\sigma_t)$ and $\xi_t$ follow an autoregressive (AR) model of order 1 (AR(1)) independently. The AR coefficients are set randomly between -1 and 1, such that all GEV parameters are stationary with respect to $t$.

\subsubsection*{Setting 2: Location parameter and scale parameter are functional}

In the second setup, we consider the location and shape parameters being function-valued, while the shape parameter is a scalar, i.e.,
\begin{align*}
\mu_t(\tau) &= \beta_{t,0} + \sum_{i=1}^{d_1}\beta^{\mu}_{t,i}b_{i}(\tau), \\
\text{ln}(\sigma_t(\tau)) & = \sigma_{t,0} + \sum_{i=1}^{d_2}\beta^{\sigma}_{t,i}b_{i}(\tau), \\
\xi_t(\tau) & = \xi_t,
\end{align*}
where $b_{i}(\cdot)$ is a cubic regression spline basis.

For the location and scale parameters, all the coefficients $\beta^{\mu}_{t,i}$ and $\beta^{\sigma}_{t,i}$ are generated from AR(1) models with respect to time $t$ independently, and we consider $d_1 = d_2 =5$. All the AR coefficients are set randomly between -1 and 1, such that the basis coefficients are stationary with respect to $t$. For the scale parameter, we generate $\xi_t$ from a stationary AR(1). 

\subsubsection*{Setting 3: All parameters are functional}

We also consider the most general setting that all GEV parameters are function-valued, that is
\begin{align*}
\mu_t(\tau) &= \beta_{t,0}^{\mu} + \sum_{i=1}^{d_1}\beta_{t,i}^{\mu}b_{i}(\tau), \\
\ln(\sigma_t(\tau)) & = \beta_{t,0}^{\sigma} + \sum_{i=1}^{d_2}\beta_{t,i}^{\sigma}b_{i}(\tau), \\
\xi_t(\tau) & = \beta_{t,0}^{\xi} + \sum_{i=1}^{d_3}\beta_{t,i}^{\xi}b_{i}(\tau),
\end{align*}
where $b_{i}(\cdot)$ is a cubic regression spline basis. Similar to setting 2, all the basis coefficients are generated from stationary AR(1) models independently, and $d_1 = d_2 = d_3 = 5$.

Once we obtained all the sampled GEV coefficients $(\mu_t(\tau_j),\sigma_t(\tau_j),\xi_t(\tau_j))$ for each grid point $\tau_j$ on the curve, where  $\tau_j\in [0,1], j=1,...,30$, we draw a random sample $\X_t(\tau_j)$ from $\text{GEV}(\mu_t(\tau_j),\sigma_t(\tau_j),\xi_t(\tau_j))$. In this simulation study, we consider $T=50$.

\subsection{Density error criteria}\label{S.error}

We measure the discrepancy between the forecast GEV density and the actual future GEV density by considering the discrete version of the JSD \citep{shannon1948mathematical} and Kullback–Leibler divergence (KLD) \citep{kullback1951information}.

The JSD measures the loss of information when we choose an approximation. For the actual and predicted probability density functions, denoted by and $f_j(\cdot)$ and $\widehat{f}_j(\cdot)$, the discrete version of the JSD is given by 
\begin{align}
    \mathrm{JSD}_j \label{JSD} &= \frac{1}{2} D_{\mathrm{KL}}\left(f_j | \delta_j\right)+ \frac{1}{2}D_{\mathrm{KL}}\left(\widehat{f}_j | \delta_j \right) \\
    & = \frac{1}{2}\sum_{i=1}^{K}f_j \left(v_{i}\right) \left[\ln \big(f_j \left(v_{i}\right)\big)-\ln \big( \delta_j \left(v_{i}\right)\big)\right]+
\frac{1}{2}\sum_{i=1}^{K} \widehat{f}_j \left(v_{i}\right)\left[\ln \big(\widehat{f}_j \left(v_{i}\right)\big)-\ln \big(\delta_j \left(v_{i}\right)\big)\right], \nonumber
\end{align}
where $\{v_i,i=1,\dots,K\}$ are the finite realisations on the density function and $\delta(\cdot)$ measures a common quantity between $f_j(\cdot)$ and $\widehat{f}_j(\cdot)$. We consider the simple mean given by $\delta_j(v) = f_j(v)+\widehat{f}_j(v)$. The JSD is locally proportional to the Fisher information metric, and is similar to the Hellinger metric, in the sense that it induces the same affine connection on a statistical manifold, and is equal to half the so-called Jeffreys divergence \citep{fuglede2004jensen}.

Alternatively, the discrepancy can be measured by the KLD, give by
\begin{align}
    \mathrm{KLD}_j \label{KLD} 
    &= D_{\mathrm{KL}}\left(f_j | \widehat{f}_j\right)+ D_{\mathrm{KL}}\left(\widehat{f}_j | f_j \right),
\end{align}
which is symmetric and non-negative.

Under the GAEV model, for each year $i$ in the forecasting period, denoted by $\widehat{\X}_i(\cdot)$ there is a unique GEV distribution for each $\tau_j$. Hence, we measure the $\text{JSD}_j^{(i)}$ and $\text{KLD}_j^{(i)}$ for each realisation $j=1,..., J$ on the curve and then take the average as the divergence measure for a testing sample $i$ in the forecasting period, that is
\begin{align*}
\text{JSD}^{(i)} &= \frac{1}{J}\sum_{j=1}^{J}\text{JSD}_j^{(i)},\\
\text{KLD}^{(i)} &= \frac{1}{J}\sum_{j=1}^{J}\text{KLD}_j^{(i)}.
\end{align*} 

\subsection{Results and Discussion}\label{S.result}

For each setting in the data generating process, we draw 250 Monte-Carlo samples to compare the forecasting performance. Within each iteration, we use an expanding window approach to measure the forecast accuracy. We use the last $20\%$ functional data as the testing sample and the rest as the training sample. We then produce iterative one-step-ahead prediction based on the three methods (see Section~\ref{S4.2} for details), where the training sample increases one data at a time. The error is measured as the mean divergence across all the test samples using the JSD and KLD. To be more specific, let $N$ be the total number of the testing data within one iteration, and $\text{JSD}^{(i)}$ and $\text{KLD}^{(i)}$ be the JSD and KLD measure for the $i^{\text{th}}$ testing data respectively. Then, the mean error for one iteration is
\begin{align*}
\text{JSD} &= \frac{1}{N}\sum_{i=1}^{N}\text{JSD}^{(i)},\\
\text{KLD} &= \frac{1}{N}\sum_{i=1}^{N}\text{KLD}^{(i)}.
\end{align*}

\begin{figure}[!htbp]
        \begin{subfigure}[b]{0.5\textwidth}
                \includegraphics[width=\linewidth]{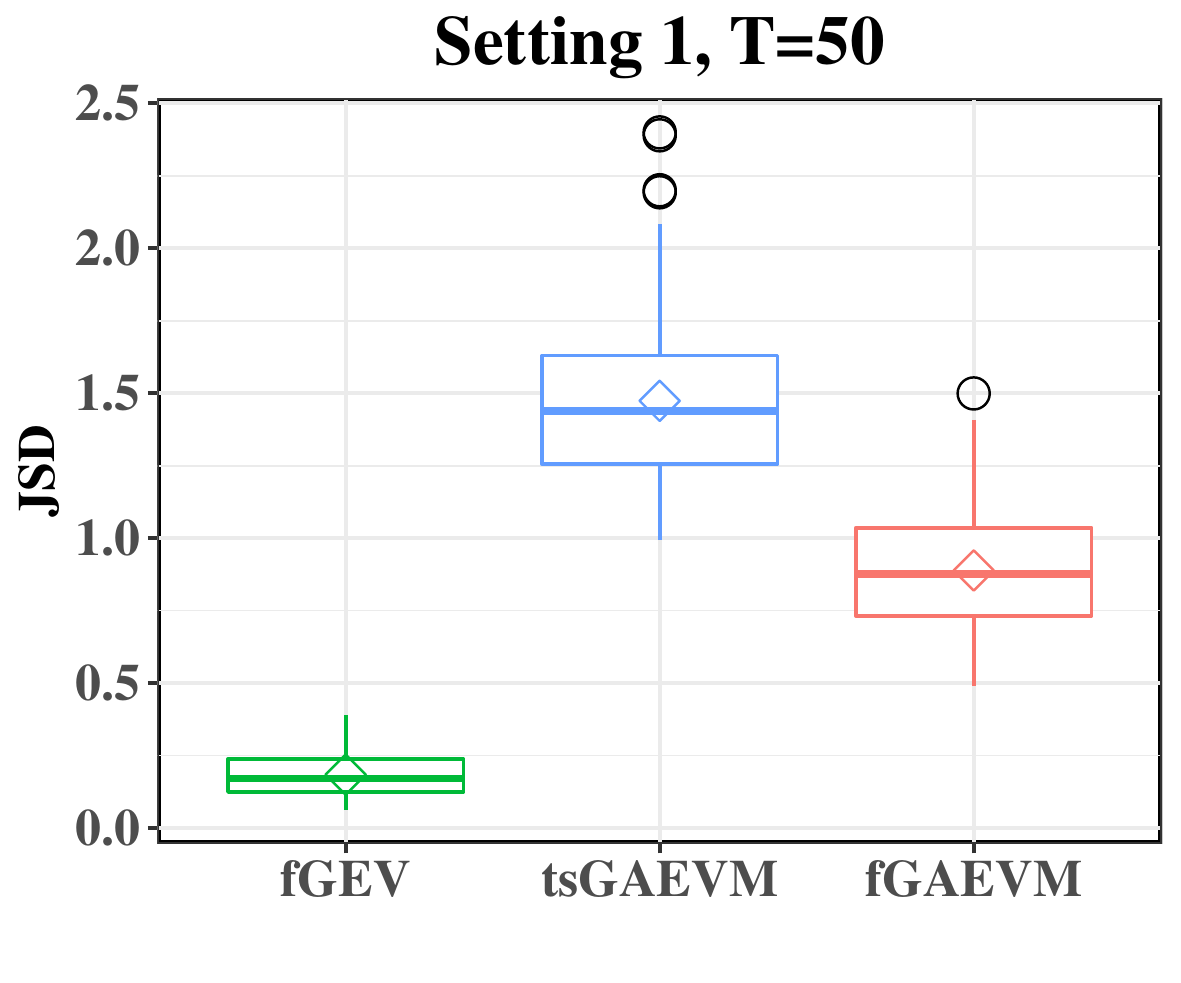}
        \end{subfigure}%
        \begin{subfigure}[b]{0.5\textwidth}
                \includegraphics[width=\linewidth]{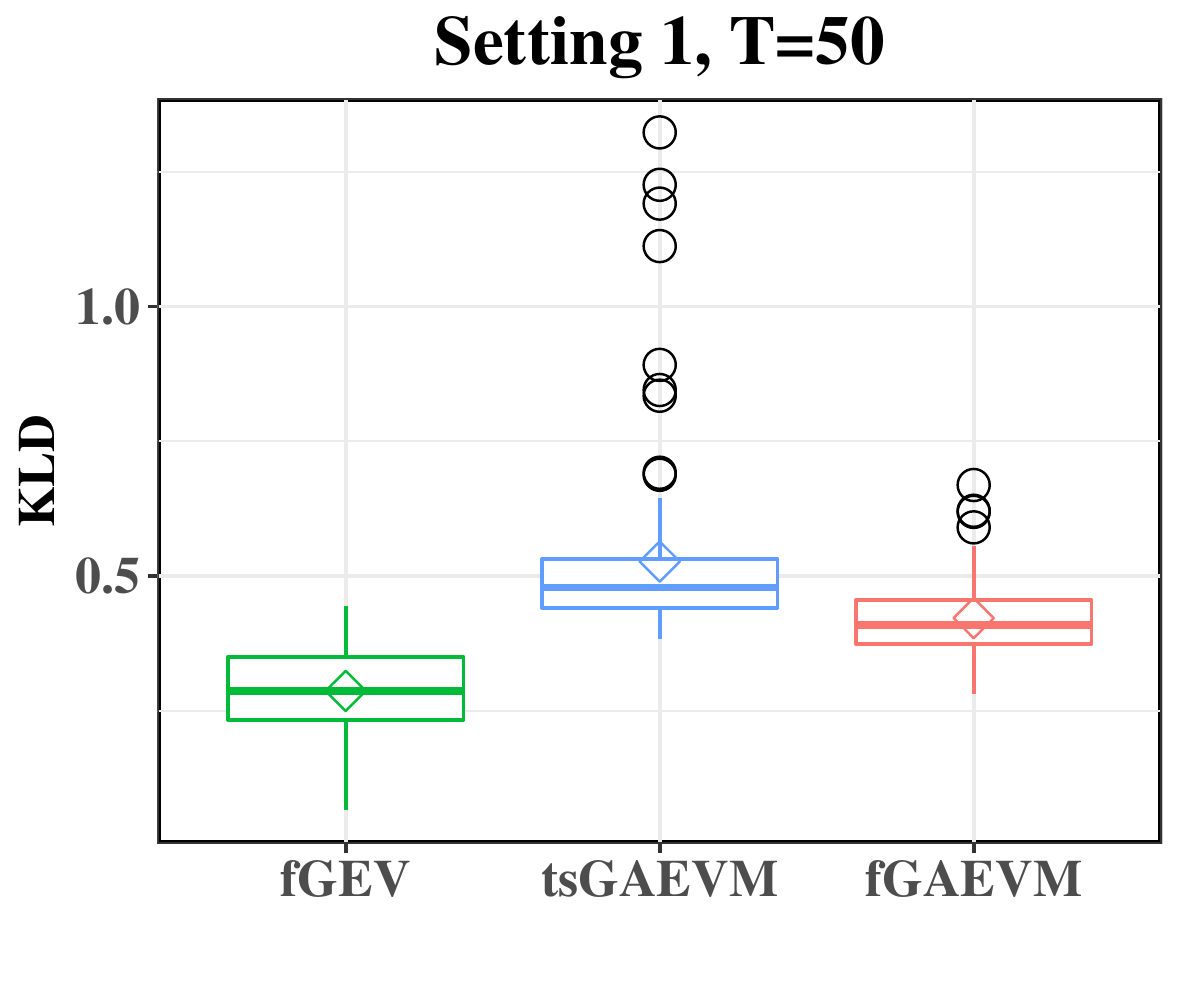}
        \end{subfigure}%
        \vfill
        \begin{subfigure}[b]{0.5\textwidth}
                \includegraphics[width=\linewidth]{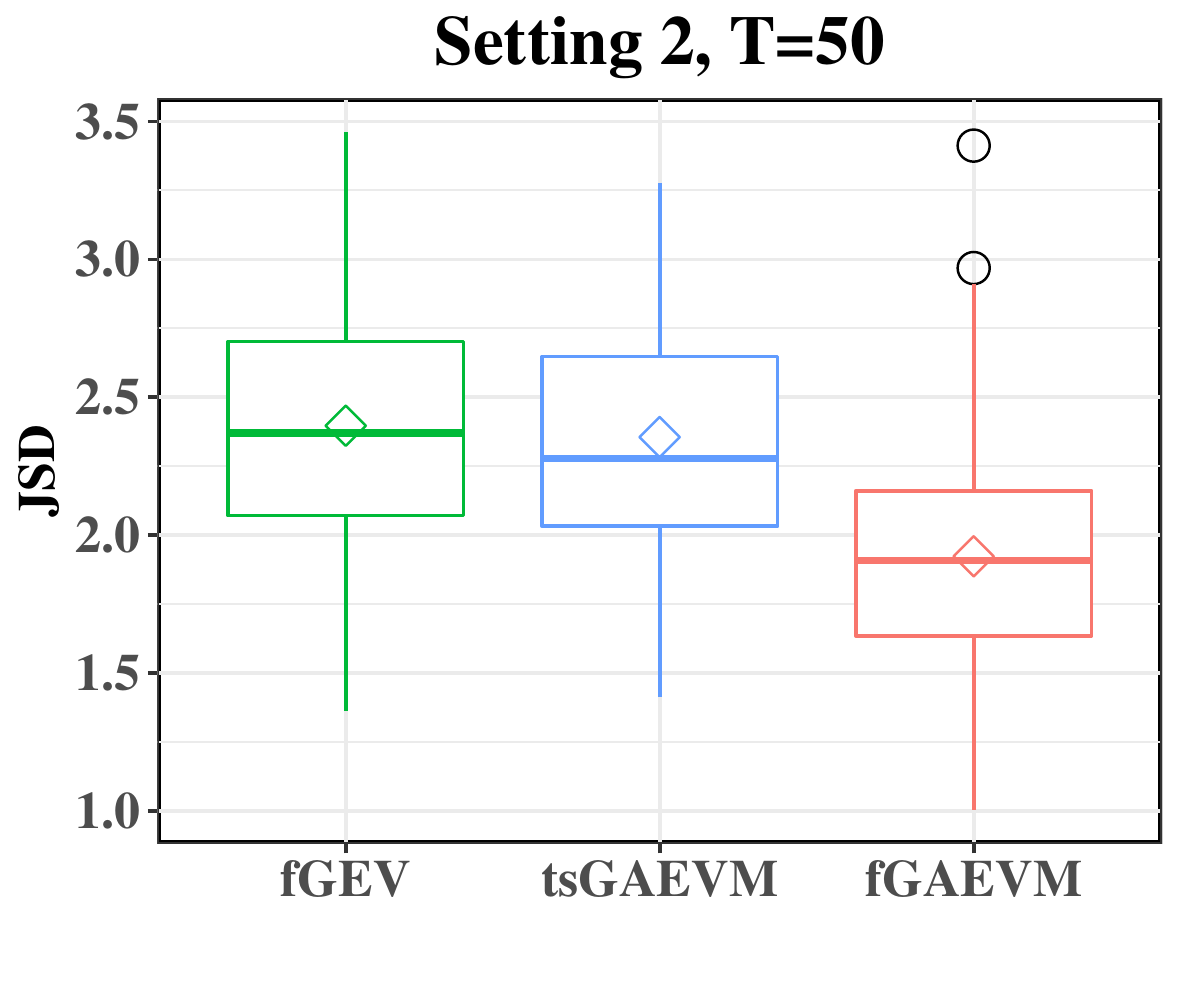}
        \end{subfigure}%
        \begin{subfigure}[b]{0.5\textwidth}
                \includegraphics[width=\linewidth]{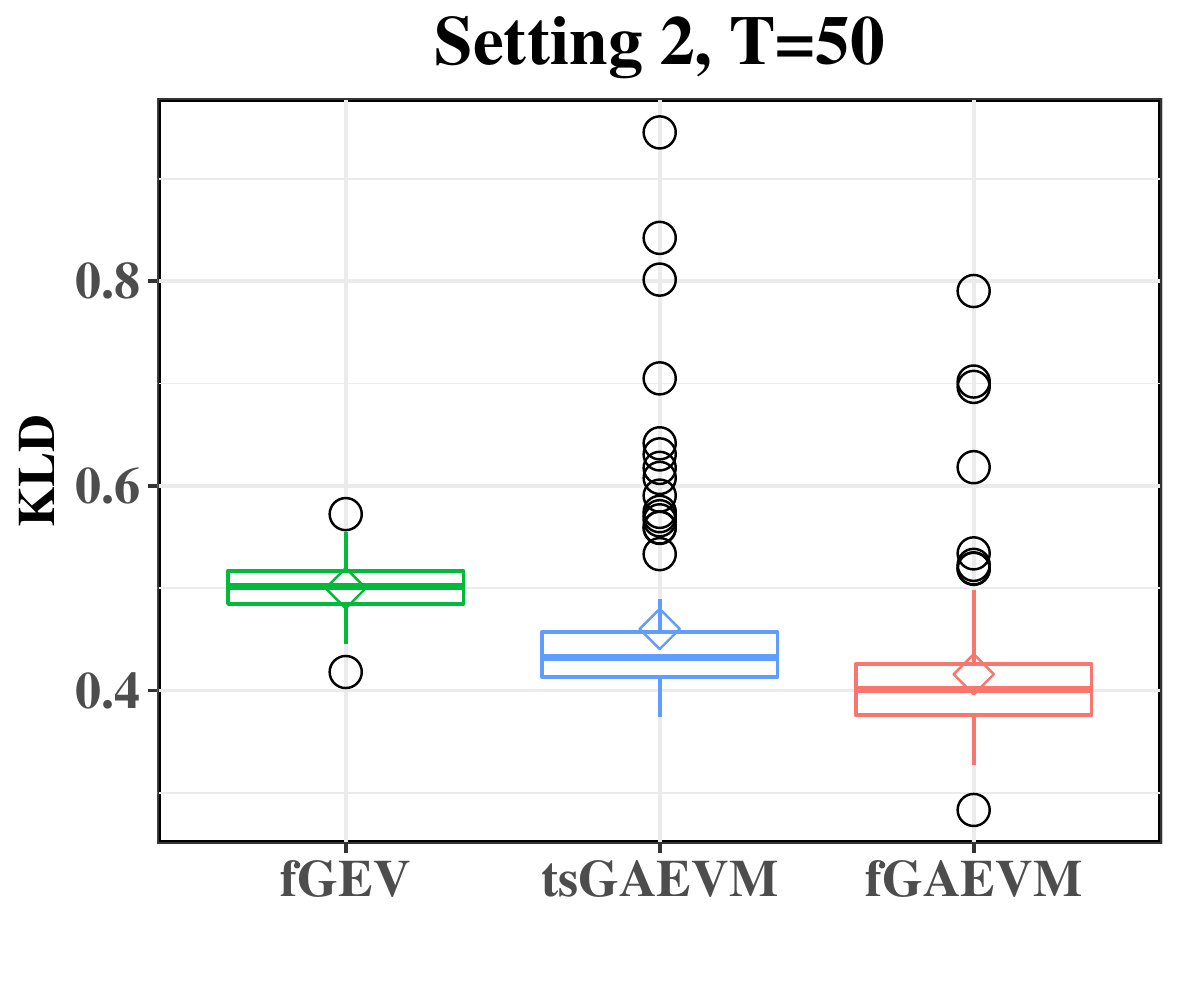}
        \end{subfigure}%
        \vfill
        \begin{subfigure}[b]{0.5\textwidth}
                \includegraphics[width=\linewidth]{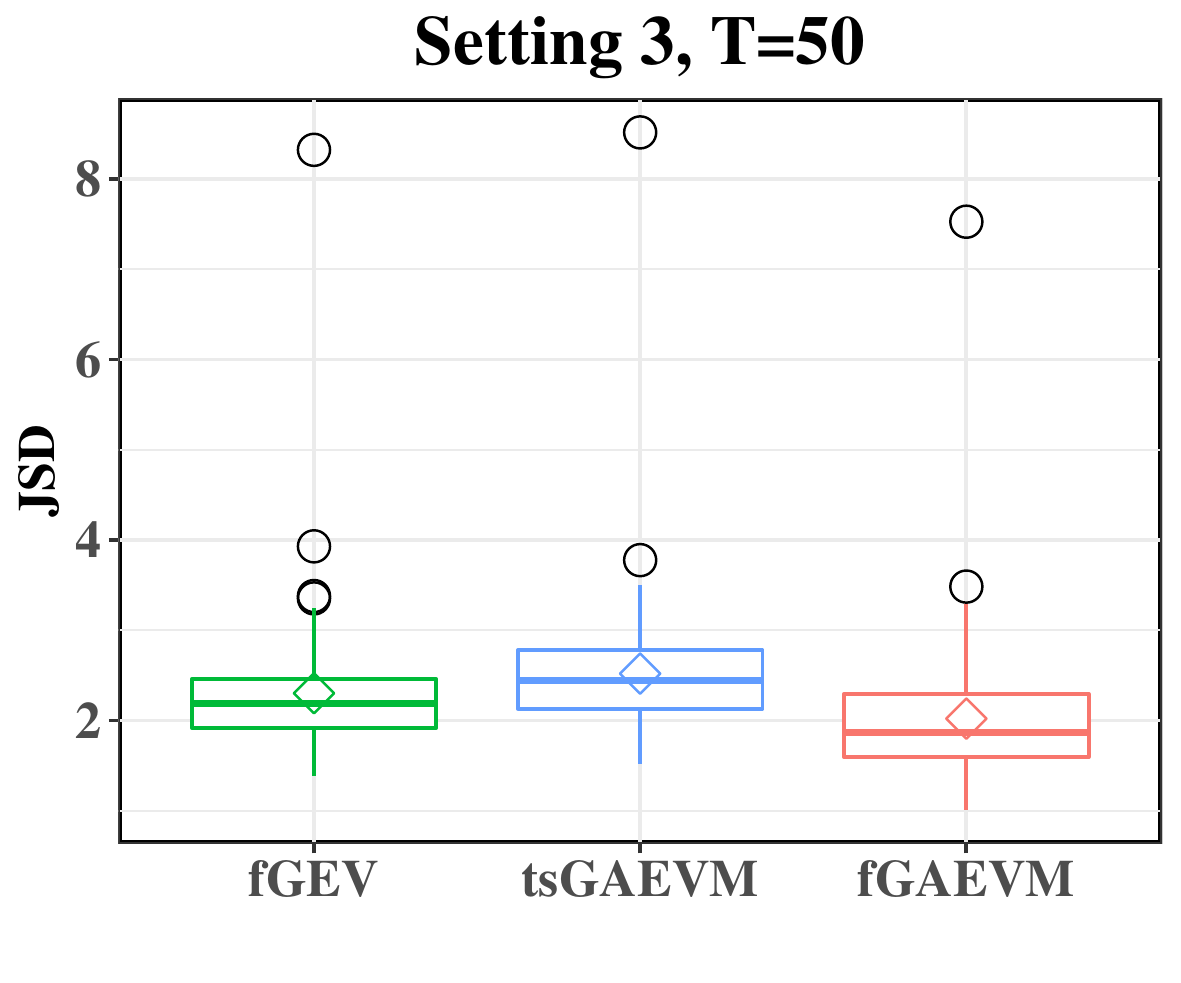}
        \end{subfigure}%
        \begin{subfigure}[b]{0.5\textwidth}
                \includegraphics[width=\linewidth]{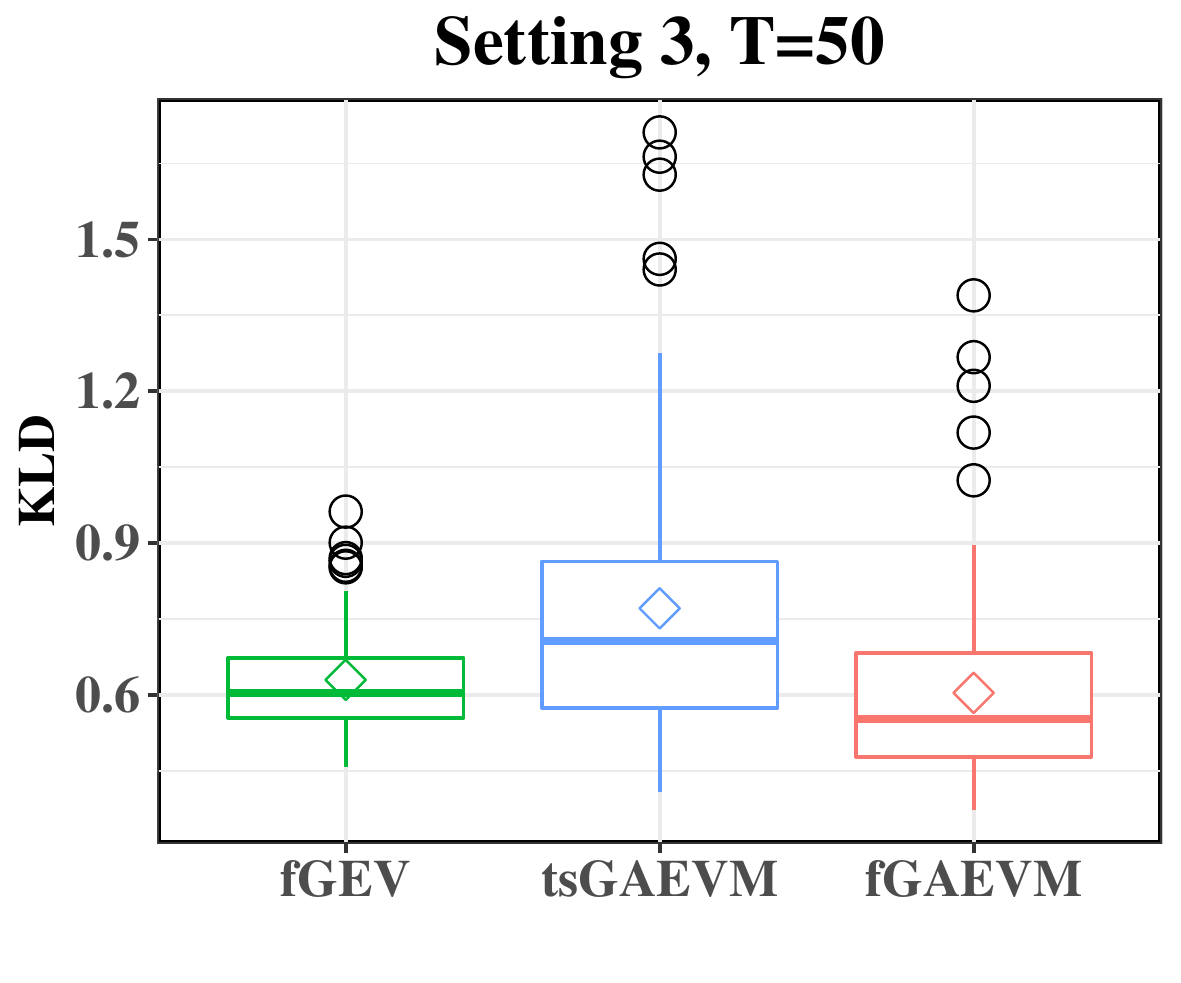}
        \end{subfigure}%
        \caption{Boxplots for the point forecast accuracy under the fGEV, tsGAEVM and fGAEVM with the simulated 250 Monte-Carlo repetitions under Setting 1, 2, and 3 for $T=50$ or $100$}
        \label{boxplot}
\end{figure}

The mean and standard deviation (in parentheses) of the averaged JSD and KLD across 250 repetitions are reported in Table~\ref{table1}. The corresponding boxplots are also presented in Figure~\ref{boxplot}.

\begin{table}[!htbp]
\centering
    \caption{Summary statistics of the averaged KLD and JSD based on the 250 simulated data\label{table1}}
\tabcolsep 0.12in
     \begin{tabular}{@{}c l  c c  c c c c @{}} 
\toprule
       &  \multicolumn{2}{c}{fGEV}& \multicolumn{2}{c}{tsGAEVM} & \multicolumn{2}{c}{fGAEVM} \\
      &  JSD & KLD & JSD & KLD & JSD & KLD\\
\midrule
      Setting 1 & \textBF{0.18(0.08)}& \textBF{0.28(0.08)}& 1.47 (0.16) & 0.82 (0.29) & 0.88(0.21) & 0.42(0.07)\\
      \\
     Setting 2  & 2.39(0.47) & 0.49(0.02) & 2.35(0.41) & 0.46(0.09)&\textBF{1.92(0.41)}&\textBF{0.41(0.07)} \\
    \\
      Setting 3 & 2.30(0.77) & 0.63(0.10) & 2.51(0.76) & 0.77(0.27)& \textBF{2.02(0.78)}& \textBF{0.60(0.18)} \\
\bottomrule
    \end{tabular}
\end{table}

Our method fGAEVM surpasses the other two methods when the data are generated from a true GAEV model (setting 2 and 3), that is a model with at least one function-valued parameter. Compared to the fGAEVM method, the fGEV fails to capture the true underlying dynamic structure in the parameters, while tsGAEVM treats the functional realisations as a univariate time series. Therefore, to obtain the one-year-ahead forecast the for the functional object $\{\widehat{\X}_{t+1}(\tau_j), j=1,\dots,J\}$, using tsGAEVM requires the forecasts from one until $J$-step, the result of which generally becomes poorer as the forecast horizon increases.

However, when data follows a GEV distribution with scalar parameters (setting~1), the fGEV model gives the best forecasting results. The fGAEVM and tsGAEVM perform on par with the fGEV, subject to all the GAM basis coefficients for the parameters are set to zero, or the number of basis coefficients equals zero. In this case, $d_1$ and $d_2$ can not be appropriately determined using the leave-one-out cross-validation with the lower bound of the parameter setting to 3 (the least degree of freedom required for a cubic regression spline). Hence, the tsGAEVM and fGAEVM model both suffer from the overfitting problem when the data follows a GEV model.

From the forecasting results in all three scenarios, we may conclude that fGAEVM gives satisfactory results in most of the scenarios, even if the true underlying nonstationary data structure is different from the model. 

Since the tsGAEVM always results in worse forecasts than the fGAEVM regardless of the underlying structure, modelling the data as a functional object is advantageous than a long univariate time series.

\section{Conclusion}\label{S5}

We proposed two general forecasting frameworks for functional time series of extreme values. One is a GEV model with scalar parameters (fGEV; see Section~\ref{S2}), and another one is a GAEV model (fGAEVM, see Section~\ref{S3}), which is a GEV model with parameters follow a GAM. As the tsGAEVM always results in worse forecasts than the fGAEVM regardless of the underlying structures (Section~\ref{S.result}), the importance of treating the data as a sliced functional time series rather than a long univariate time series has been revealed. Illustrated by the maximum temperature data in Sydney in Section~\ref{S4.4} and simulated data in Section~\ref{S.result}, when the underlying data follows a GAEV model, the fGEV gives a senseless result of a horizontal line along with the functional object. At the same time, the fGAEVM can capture the dynamics along with the functional object. On the other hand, if the data indeed follows a static GEV distribution, the large flexibility of the fGAEVM may reduce the forecast accuracy compared to the fGEV model. Since the static GEV distribution and the GAEV distribution can be distinguished based on prior knowledge of the data set or using data visualisations, we can choose between the fGEV and fGAEVM adaptively. 

There are a few ways in which this paper could be extended, and we briefly discuss three. Firstly, We can model the temporal dependency along day-in-year in a more precise way. In this study, we consider daily maxima. Instead of block maxima, we may also consider identifying ``independent'' clusters and then fitting a GEV model to cluster maxima or fit the GEV under a Bayesian framework. Secondly, we use the cubic regression spline basis to represent the generalised additive model for the GEV parameters. In future research, other bases may be utilised, and comparison can be made. Finally, in the empirical analysis, the forecasting uncertainty is assessed using a sieve bootstrapped prediction interval method proposed for functional time series \citep{Efstathios2020pi}. Bootstrap method tailored for the fGAEVM density may be a promising future research topic.



\newpage
\bibliographystyle{agsm}
\bibliography{GEV_functional.bib}

\end{document}